# Spontaneous emergence of Josephson junctions in homogeneous rings of single-crystal Sr$_2$RuO$_4$

Yuuki Yasui[1†**], Kaveh Lahabi[2†], Victor Fernández Becerra[3,4], Remko Fermin[2], Muhammad Shahbaz Anwar[1,5], Shingo Yonezawa[1], Takahito Terashima[1], Milorad V. Milošević[3], Jan Aarts[2] & Yoshiteru Maeno [1*]

[1]Department of Physics, Graduate School of Science, Kyoto University, Kyoto 606-8502, Japan.

[2]Huygens – Kamerlingh Onnes Laboratory, Leiden Institute of Physics, Leiden University, P.O. Box 9504, 2300 RA Leiden, The Netherlands.

[3]Department of Physics, University of Antwerp, Groenenborgerlaan 171, B-2020 Antwerp, Belgium

[4]International Research Centre Magtop, Institute of Physics, Polish Academy of Sciences, Aleja Lotników 32/46, PL-02668 Warsaw, Poland.

[5]Department of Materials Science and Metallurgy, University of Cambridge, 27 Charles Babbage Road, Cambridge CB3 0FS, UK.

[†]These authors contributed equally to this work.

* Correspondence to YM (maeno@scphys.kyoto-u.ac.jp)

** Present address: RIKEN Center for Emergent Matter Science, Wako, Saitama 351-0198, Japan.

**Abstract**

The chiral *p*-wave order parameter in $Sr_2RuO_4$ would make it a special case amongst the unconventional superconductors. A consequence of this symmetry is the possible existence of superconducting domains of opposite chirality. At the boundary of such domains, the locally supressed condensate can produce an intrinsic Josephson junction. Here, we provide evidence of such junctions using mesoscopic rings, structured from $Sr_2RuO_4$ single crystals. Our order parameter simulations predict such rings to host stable domain walls across their arms. This is verified with transport experiments on loops, with a sharp transition at 1.5 K, which show distinct critical current oscillations with periodicity corresponding to the flux quantum. In contrast, loops with broadened transitions at around 3 K are void of such junctions and show standard Little-Parks oscillations. Our analysis demonstrates the junctions are of intrinsic origin and makes a compelling case for the existence of superconducting domains.

**Introduction**

$Sr_2RuO_4$ stands out among the unconventional superconductors as one of the few materials with a chiral order parameter [1,2]. The tetragonal crystal structure allows five unitary representations for a *p*-wave pairing symmetry [1,3]. One of these is the chiral order parameter, of the form $k_x \pm ik_y$, which is strongly suggested by muon spin relaxation [4] and high-resolution polar Kerr effect measurements [5]. Very recently, nuclear magnetic resonance experiments demonstrated that the *d*-vector is not parallel to the *c*-axis and suggested possible chiral *d*-wave states [40,41]. Such chiral states are attracting renewed attention due to the possibility of hosting Majorana bound states, which in turn are of interest for topological quantum computing [6,7,8]. A key property of the chiral state is its double degeneracy in the orbital degree of freedom, with important consequences such as the existence of superconducting domains of different chirality and a spontaneous edge current. The major problem plaguing our understanding of $Sr_2RuO_4$ [9] is that, although the chiral state seems probable, domains or edge currents have not been observed directly. Indications for their existence, however, have been found in transport experiments, which utilize Ru inclusions to form proximity junctions between Sr$_2$RuO$_4$ and a conventional *s*-wave superconductor [10,11]. A complication in the physics of Sr$_2$RuO$_4$ is that breaking of the tetragonal crystal symmetry due to Ru inclusions or a uniaxial strain can induce a different superconducting state with an enhanced superconducting transition temperature $T_\text{c} \approx$ 3 K [13,14]. Recent experiments suggest that this so-called 3-K phase may exhibit a non-chiral state with a single-component order parameter [15,16]. In this paper, we refer to the multi-component phase with $T_\text{c}$ of around 1.5 K, associated with the pure bulk limit, as the "intrinsic phase" and the possible single-component phase, characterized by $T_\text{c} \approx$ 3 K, as the "extrinsic phase".

The vast majority of experiments in the past two decades have been limited to bulk crystals, typically hundreds of microns in dimension. This is partly due to the unavailability of superconducting Sr$_2$RuO$_4$ films. The chiral domains, however, are expected to be no more than a few microns in size [5,11]. Moreover, the time-dependent switching noise observed in transport measurements suggests the domains are mobile [10,11]. We note here that the role of chiral domains resulting in hysteretic behaviour has been discussed in the Bi-Ni bilayer system [12]. The arbitrary configuration of the domains introduces an element of uncertainty. On the other hand, the energy cost associated with a chiral domain wall (ChDW), grows per area [17]. It has been recently discussed that mesoscopic samples made of chiral *p*-wave superconductors could host multichiral states [18,19], where the two $k_x \pm ik_y$ chiral components are divided into superconducting domains, separated by ChDWs. This makes mesoscopic structures a promising platform to verify and potentially control the domains.

Another interesting aspect of a ChDW is that it can act as a Josephson junction [17] due to the local suppression of the order parameter, as schematically shown in Fig. 1a.

Here, we present results of transport measurements on mesoscopic rings of Sr$_2$RuO$_4$, prepared by focused ion beam (FIB) milling of single crystals. Homogeneous structures, characterized by a sharp transition at around the intrinsic $T_c$ of 1.5 K, show distinct critical current oscillations — similar to that of the classical DC superconducting quantum interference device (SQUID), consisting of two artificially prepared Josephson junctions. Despite the absence of conventional weak links, the interference pattern appears over the full temperature range below $T_c$ while maintaining its overall shape. In contrast, the SQUID oscillations are entirely absent in rings that are in the extrinsic phase. These systems behave as standard superconducting loops: they exhibit the conventional Little-Parks (LP) $T_c$ oscillations, which can only be observed near the resistive transition [19]. We also present calculations on the possible chiral-domain configurations for a *p*-wave superconducting ring, using the Ginzburg-Landau (GL) formalism. Experiments and calculations together make a convincing case for the existence of ChDWs in the intrinsic phase of Sr$_2$RuO$_4$.

**Results**

**Basic transport properties of single crystal microrings.** Single crystals of Sr$_2$RuO$_4$ were grown with the floating zone method [21] and structured into microrings using Ga-based FIB etching. Figs. 1b-d show scanning electron microscope (SEM) images of Rings A and B. The inner and outer radii of Ring A are $r_{\text{in}} \approx 0.21$ µm and $r_{\text{out}} = 0.55$ µm, respectively. Similar dimensions are used in Ring B; $r_{\text{in}} = 0.3$ µm and $r_{\text{out}} = 0.54$ µm. Both crystals have a thickness of around 0.7 µm.

The temperature-dependent resistance $R(T)$ of both rings (presented in Figs. 2a and b), show sharp superconducting transitions similar to that of bulk Sr$_2$RuO$_4$. The apparent enhancement of the resistance just above $T_c$ in Fig. 2b could be attributed to changes in the current path [39]. The high quality of the sample is also evident by their particularly high residual resistivity ratio; RRR = $R(300\text{ K})/R(3\text{ K})$ = 238 for Ring A and RRR = 177 for Ring B. To demonstrate that FIB milling does not alter the intrinsic characteristics of Sr$_2$RuO$_4$, we compare the $R(T)$ of Ring A with the one measured before milling the crystal in Supplementary Figure 1, which shows that $T_c$ and the overall transport properties remain unchanged under structuring. Fig. 2c shows the typical current-voltage $V(I)$ behaviour at different temperatures. For both rings, the $V(I)$ measurements exhibit negligibly small hysteresis even at temperatures far below $T_c$.

**Insights from theoretical simulations.** Before presenting the results of transport measurements under a magnetic field, we examine the expected chiral-domain configurations in our structure. This is accomplished by performing detailed time-dependent GL simulations, under the assumption of a chiral *p*-wave order parameter, for microrings with nanostructured transport leads (similar to the one used in our experiments). The simulations show that the ring can host a mono-chiral-domain or a multi domain states, depending on the parameters $\frac{r_{\text{in}}}{\xi(T)}$ and $\frac{r_{\text{out}}}{\xi(T)}$, which correspond to the inner and outer radii of the ring, scaled by the temperature-dependent coherence length $\xi(T) = \xi(T=0)\frac{\sqrt{1-t^4}}{1-t^2}$ [22, 23], where $t = \frac{T}{T_c}$, with $T_c \approx 1.75$ K for Ring A and $T_c \approx 1.3$ K for Ring B (shown in Figs. 2a and b). Based on our critical field measurements we estimate $\xi(T=0) \sim 66$ nm, which is the same as the bulk value for Sr$_2$RuO$_4$. Figure 3a shows the simulated Cooper-pair density $|\Psi|^2$ of Ring A far below $T_c$, obtained by setting $\frac{r_{\text{in}}}{\xi(T)} = 2.5$ and $\frac{r_{\text{out}}}{\xi(T)} = 6.8$ (corresponding to $T \lesssim 0.5\, T_c$ in our measurements). This state contains two distinct chiral domains, separated by a pair of ChDW. Within the domain wall the order parameter is reduced to about half of its original amplitude in the banks on each side, resulting in the formation of two parallel Josephson weak links. While the suppressed order parameter is unfavourable in terms of the condensation energy, the formation of such ChDW is favoured by the second term of the free energy in the Supplementary Eq. (S11). Since the order parameter is suppressed at the sample edge, the second term gains importance with reducing sample size and may further enhance an inhomogeneous order-parameter state with a ChDW. The ChDW region extends over a length of the order of $\xi$. As shown in Fig. 3b, the presence of a magnetic field along the ring axis makes the positions of the ChDWs shift away from the middle of the arms since one of the chiral components is favoured by the magnetic field. The ChDWs, however, remain in the arms of the ring due to the strong pinning by the restricted dimensions.

Figure 3c shows the calculated chiral-domain configuration for Ring B, which also applies to Ring A at temperatures near $T_c$. This is obtained by setting $\frac{r_{\text{in}}}{\xi(T)} = 1.3$ and $\frac{r_{\text{out}}}{\xi(T)} = 3.6$ (corresponding to $T \approx 1.45$ K for Ring A). As the arms of the ring are now considerably narrower on the scale of $\xi(T)$, the contribution of the edge regions dominates the configuration of the order parameter. As a consequence, it becomes energetically favourable for the two chiral components to coexist over the entire ring. This state also produces a pair of parallel weak links due to the suppression of the order parameter $|\Psi|$, which extend over the arms of the rings. Fig. 3d presents a phase diagram of the lowest energy states, calculated for various $\left(\frac{r_{\text{in}}}{\xi(T)}, \frac{r_{\text{out}}}{\xi(T)}\right)$. Amongst these, mono-domain "Meissner" state can be stabilised by increasing the $r_{\text{out}}/r_{\text{in}}$ ratio. In this state, the arms of the ring are unable to provide effective pinning of ChDWs. This scenario is explored in Supplementary Note 4 and Supplementary Figure 5, where a ring with relatively wide arms (Ring E) approaches the mono-domain state at low temperatures. The evolution of the equilibrium domain configuration as a function of temperature for Rings A and B are represented by the dashed lines in Fig. 3d. This suggests that the rings are in one of the domain states shown in Figs. 3a and c at all temperatures below $T_c$, except in a narrow range around 1 - 1.2 K, where additional domain walls could appear in Ring A. As a general finding, our GL calculations show that ChDWs could spontaneously emerge in our mesoscopic rings and behave as stable Josephson junctions over a broad temperature range, resulting in a DC SQUID of intrinsic origin. The change of chirality across such junctions and its influence on their transport characteristics remain open questions and are worthy of further studies. Note that the GL formalisms for chiral *p*-wave and chiral *d*-wave superconductors have analogous form, and the segregation of chiral domains as discussed above is applicable to both cases.

**Critical current oscillations.** We examined the supercurrent interference of the rings by measuring $I_c$ at each magnetic field $H$. The results are presented in Fig. 4, where we observe the same behaviour in both Rings A and B. Figures 4a and b show the $I_c$ of Ring A, measured for positive ($I_{c+}$) and at negative ($I_{c-}$) bias currents, taken at temperatures deep inside the superconducting state and close to $T_c$, respectively. For both temperatures, we observe distinct critical current oscillations, with the period corresponding to the fluxoid quantization over the ring area. This interference pattern corresponds to that of a DC SQUID with a pair of parallel Josephson junctions. The junctions would also need to be symmetric each other; an imbalance in $I_c$ could not produce the cusp-shaped minima of the patterns. The figure also shows $-I_{c-}(-H)$ overlaid on its time-reversed counterpart, $+I_{c+}(H)$. Figure 4c shows that the same SQUID oscillations appear in Ring B, only with a slightly smaller period (consistent with its slightly larger inner radius). The oscillations emerge spontaneously at the onset of superconductivity and continue down to $T \ll T_c$. More importantly, we find that the patterns are not distorted, despite the substantial variations in $I_c(T)$ and $\xi(T)$.

It is worth noting that, unlike the polar Kerr experiments, we find field cooling and zero-field cooling of the samples to yield the same results in our measurements. This, however, is to be expected in mesoscopic structures, where domain walls are strongly pinned to the confined regions in order to lower the free energy of the system (see Supplementary Note 2 and Supplementary Figure 2 for more details). Such pinning mechanism is absent in the polar Kerr experiments, which are performed on bulk crystals [5].

To demonstrate the robustness of the SQUID behaviour further, in Figs. 5a and b, we plot the magnetoresistance of Ring A, produced by the $I_c$ oscillations over a wide range of temperatures. These are measured by applying a constant DC current $\pm I$ while sweeping the magnetic field $H$ along the ring axis. Here, the resistance $R$ is defined by the average of two voltages before and after current reversal at each magnetic field; $R = [V(I) - V(-I)]/2I$. When the measurement current exceeds the critical current $I_c(H)$, the system is driven out of the zero-voltage regime of the $V(I)$ and produces a finite resistance. Combining the results of a wide range of temperatures, Figs. 5a and b reveal that the SQUID oscillations emerge together with $I_c$ at the onset of the superconducting transition. In Figs. 5c and d, we describe the shape of $R(H)$, where in some cases the peaks can appear to be split or broadened. This is clearly due to a slight difference in the values of $I_{c\pm}$, which causes the voltage peaks for $\pm I$ to appear asymmetrically. We observed a similar asymmetry in rings showing the LP effect [20].

The magnetovoltage and field-dependent $V(I)$ measurements are crucial in resolving an outstanding issue regarding previous reports of unconventional behaviour of Sr$_2$RuO$_4$ rings. Cai *et al.* have consistently observed magnetoresistance oscillations with unexpectedly large amplitude [25,26], very similar to the data presented in Fig. 5. The reported magnetoresistance oscillations are also stable over a wide range of temperatures and, in some cases, show small dips around $\Phi_0/2$. As Fig. 5 demonstrates, however, the averaged resistance $R$ could produce a very similar effect even when there is no splitting of the peaks in the raw magnetovoltage signal.

$T_c$ **oscillations in rings with an extrinsic phase.** We already mentioned that ChDWs can produce the observed $I_c(H)$ oscillations by acting as Josephson junctions. This should be contrasted with the fluxoid-periodic behaviour of structures with a partial or full extrinsic phase, characterised by a noticeably broader transition which begins near 3 K (see Fig. 6e and Supplementary Figure 4). We recently reported observations of the LP oscillations in such Sr$_2$RuO$_4$ microrings [20], and here we demonstrate that those are of a fundamentally different nature than the $I_c$ oscillations discussed in this report. For this, we compare the data from Ring A with those of Ring C (Sample B in ref. [20]), where the transition is considerably broader (Fig. 6e). This ring was prepared from a 2-μm thick crystal with a $T_c$ of 1.5 K. After microstructuring, however, the ring was found to have a higher $T_c$, with its transition already starting at 2.7 K. The magnetotransport measurements reveal that the ring itself is predominantly in the extrinsic phase, introduced by microstructuring (most likely due to a strain induced by FIB milling of the thick crystal). Compared to Rings A (RRR = 238) and B (RRR = 177), this structure has a smaller residual resistivity ratio RRR = 129. Nevertheless, the value of RRR is still substantial, indicating strong metallicity for Ring C. Figures 6a and b show $R(H)$ for temperatures within the resistive-transitions of Rings A and C (taken 1.67 K and 2.3 K, respectively). In both cases we find fluxoid-periodic oscillations, which we compare with simulated LP oscillations (the red curves).

The change of the transition temperature due to the LP oscillations is given by [28]:

$$\frac{T_c(H) - T_c(0)}{T_c(0)} = -\left(\frac{\pi\xi(0)w\mu_0 H}{\sqrt{3}\Phi_0}\right)^2 - \frac{\xi^2(0)}{r_{in}r_{out}}\left(n - \frac{\pi\mu_0 H r_{in}r_{out}}{\Phi_0}\right)^2, \quad (1)$$

where $\Phi_0 = h/2e$ is the flux quantum with the Planck constant $h$ and the elementary charge $e$, and $w = r_{out} - r_{in}$ is the width of a ring arm. The first term represents the effect of the Meissner shielding, and the second term corresponds to fluxoid quantization. To convert the change of the transition temperature to the resistance variation, we assume that the $R(T)$ curve does not change its shape under magnetic field and shifts horizontally by $\Delta T_c(H) = T_c(H) - T_c(0)$. For the simulations in Figs. 6a and b, we used $\xi(0) = 66$ nm, $2r_{in} = 0.55$ µm, $2r_{out} = 1.1$ µm for Ring A, and $2r_{in} = 0.7$ µm, $2r_{out} = 1.0$ µm for Ring C. Both the period and amplitude of the oscillations for Ring C agree with those of the simulation. We therefore consider these to be the LP oscillations, driven by variations in $T_c$. For Ring A, however, the oscillation amplitude is substantially larger than what $T_c$ variations can produce. Such large-amplitude magnetoresistance is driven by the $I_c(H)$ oscillations instead. In Figs. 6c and d, we compare the $I_c(H)$ of both rings at lower temperatures. In contrast to Rings A and B, the SQUID oscillations are completely absent in Ring C. Instead, for all temperature below $T_\text{c}$, we only observe a monotonous decay of $I_\text{c}(H)$. We find the lack of Josephson junctions to be a common characteristic among structures with a dominant extrinsic phase. A further example of this is given in Ring D (Supplementary Note 3 and Supplementary Figure 4).

**Discussion**

Before adopting ChDW scenario as the origin of the observed $I_\text{c}$ oscillations, we consider other known mechanisms for $I_\text{c}$ oscillations. Firstly, even in a homogeneous loop SQUID-like behaviour may emerge depending on the size of the ring with respective to either the penetration depth $\lambda$ or the coherence length $\xi$. $I_\text{c}$ can be modulated by the circulating persistent current $I_\text{p}$, which varies linearly with the flux, and switches its direction at every increment of $\Phi_0/2$. This mostly results in a sawtooth-like modulation of $I_\text{c}$ [31], which cannot account for non-linear form of the patterns shown in Fig. 4. Furthermore, the magnitude of $I_\text{p}$ is inversely proportional to the kinetic inductance $L_\text{K}$, which depends on the penetration depth $L_\text{K} \propto \lambda^2(T)$. If the $I_\text{c}$ oscillations were driven by circulating currents, their amplitude $\Delta I_\text{c}$ would grow larger by lowering the temperature since $\Delta I_\text{c} \propto I_\text{p} \propto 1/\lambda^2(T)$ [31]. This is clearly not the case for the Sr$_2$RuO$_4$ rings, where oscillation amplitude is unaffected by temperature (*e.g.* $\Delta I_\text{c} \approx 12$ µA at both temperatures shown in Fig. 4a). SQUID oscillations can also emerge in loops without weak link, if the dimensions are much smaller than $\xi(T)$ and $\lambda(T)$ [32]. However, this is not applicable to our structures, where the radii and the width of the arms are several times larger than the characteristic length scales for $T \ll T_\text{c}$ (*e.g.* for Ring A, $\xi(T) \sim 0.07$ µm and $\lambda(T) \sim 0.19$ µm at $T = 0.78$ K).

Secondly, Cai *et al.* attributed the large-amplitude magnetoresistance of their Sr$_2$RuO$_4$ rings to current-excited moving vortices [25, 26]. As demonstrated by Berdiyorov *et al.* [29], this mechanism can only produce large-amplitude oscillations over a finite temperature range, typically down to $T \sim 0.95\, T_\text{c}$ (*e.g.* see Fig. 6b of [29] and Fig. 2 of [30]). This is not the case for the Sr$_2$RuO$_4$ rings that are in the intrinsic (1.5-K) phase, as the magnetoresistance oscillations appear for all $T < T_\text{c}$ (see Figs. 5a, b and Fig. 3a in [25]).

Thirdly, geometrical constrictions (*e.g.* bridges and nanowires) can serve as Josephson junctions, as long as their dimensions are comparable to $\xi$. The current-phase relation (CPR) of such junctions is defined by the ratio of $\xi(T)$ to the length of the weak link $L$. Since $\xi(T)$ varies with temperature while $L$ remains fixed, the CPR of such weak links is strongly temperature dependent. Generally, lowering the temperature transforms the CPR from sinusoidal to a sawtooth-like function, which ultimately turns into multivalued relations once $L \gtrsim 3.5\, \xi(T)$, corresponding to the nucleation of phase-slip centres [33,34,35]. The multivalued CPR manifests itself as a hysteretic $V(I)$ relation, which is a well-known characteristic of constriction junctions at $T \ll T_\text{c}$ [36,37]. This is in direct contrast to the $V(I)$ curves of the Sr$_2$RuO$_4$ rings, which show negligible hysteresis for temperatures as low as $0.2\, T_\text{c}$ (see Fig. 2c). Furthermore, the interference patterns taken at over wide range of temperatures show the same overall shape, with characteristically round lobes (Fig. 4). This could not be produced by constriction junctions, as the interference pattern would be heavily deformed by the pronounced changes in $\xi(T)/L$ with temperature. In case of ChDWs, however, the length of the junction barrier is determined by the coherence length and therefore has a temperature dependence similar to $\xi(T)$. Hence, a ChDW junction can maintain a relatively fixed $\xi(T)/L(T)$ ratio for different temperatures. This would agree with the lack of hysteresis in our $V(I)$ measurements (Fig. 2c) and the unperturbed shape of the interference patterns (Fig. 4).

Lastly, we exclude the possibility of forming accidental proximity junctions by Ru inclusions or any other normal metal within the Sr$_2$RuO$_4$ crystal. Apart from their absence in the SEM images taken while the milling of the rings, inclusions would induce an extrinsic 3-K phase. The crystals, however, show no such enhancement of $T_c$ either before or after FIB processing. Moreover, the (single) sharp resistive transitions of Rings A and B could not be produced in presence of normal metal weak links. Accidental tunnel junctions, formed by nanocracks or grain boundaries, can also be excluded due to the high metallicity of our samples. In summary, the Josephson effect found in Sr$_2$RuO$_4$ microrings cannot be attributed to conventional types of weak link such as constriction junctions, kinematic vortices (phase-slip lines), proximity and tunnel junctions.

To summarise, our simulations of a chiral *p*-wave order parameter show that a mesoscopic loop with nanostructured transport leads can host a multi-domain state. The degenerate chiral states are separated by ChDWs located in the arms of the ring, where a pair of parallel Josephson junctions is formed due to the local suppression of both chiral states. We examined the existence of such junctions by performing transport experiments on Sr$_2$RuO$_4$ microrings. The rings with a sharp transition near 1.5 K show distinct $I_c$ oscillations, similar to that of a DC SQUID with a pair of Josephson junctions with matching $I_c$. The junctions emerge together with the superconducting transition and are present for all temperatures below $T_c$. In contrast, for Sr$_2$RuO$_4$ rings with an extrinsic (3-K) phase, the Josephson junctions are entirely absent. Such rings show standard Little-Parks oscillations near $T_c$, which can be properly modelled, but no critical current oscillations. Our findings suggest that the Josephson junctions are an inherent property of the order parameter, and make a compelling case for the existence of ChDWs in the intrinsic (1.5-K) phase of $Sr_2RuO_4$. We should note that our present results formally do not distinguish the type degenerate states responsible for the formation of the junctions; our transport measurements would also be consistent with domain walls of helical states, as well as of spin-singlet chiral states. This work also demonstrates that the combination order parameter simulations with mesoscopic structures can be instrumental in the study of superconducting domains and will, in coming experiments, allow for detailed design and understanding of a system before the actual fabrication.

**Methods**

**Microring fabrication**. $Sr_2RuO_4$ single crystals were prepared with the floating zone method [21], and their transition temperature $T_c$ before the sample fabrication was confirmed to be 1.50 K using a compact AC susceptometer [38] in a Quantum Design PPMS. We crush the crystal into small pieces to obtain thin crystals with the thickness of approximately 1 μm. Although $Sr_2RuO_4$ is chemically stable in the ambient condition, we find that small crystals can degrade in the air. Therefore, freshly crushed crystals were used. The crystal is placed on a $SrTiO_3$ substrate, where it is contacted by either gold or silver for transport measurements. For Rings A, C and D, two pads of high-temperature-cure silver paint (6838, Dupont) are attached to the two sides of the crystal. The paint is then cured at 500°C for 20 minutes. In case of Rings B and E however, the crystals are contacted using a combination of electron-beam lithography and sputter deposition of gold. Once a crystal is contacted by the gold or silver paint, a 100-nm thick layer of $SiO_2$ is deposited using electron beam evaporation to protect the crystal during structuring. The contacts and the crystal underneath are then cut with a Gallium focused ion beam (FIB) to produce a four-wire arrangement. Lastly, the microrings are structured using the FIB (30 kV, 20 pA).

**Measurements**. Transport measurements were performed in a ³He refrigerator (Heliox, Oxford Instruments) down to 0.3 K. In the DC resistance measurement, we flip the direction of the measurement current to subtract the contribution of the thermoelectric voltage, and the resistance $R$ is defined to be $R = [V(I) - V(-I)]/2I$. The transition temperature shift due to the LP oscillations is calculated to be approximately 10 mK by using Eq. (1). Therefore, temperature stability the during magnetoresistance measurement must be much smaller than this value. By putting a 80-Ω by-pass resister in parallel to the heater and by tuning the PID values of the temperature controller, we achieved a temperature stability of 100 μK. Current-voltage $V(I)$ measurements are performed under constant temperature and magnetic field with triangular current waves of frequency 2 mHz.

**Simulations.** For details of the Ginzburg-Landau simulations we refer to the formalism of [18], and the additional discussion in the Supplementary information.

## Data availability

The data that support the findings of this study are available from the corresponding author upon reasonable request.

## Acknowledgements

The authors would like to thank S. Goswami, A. Singh, M. Kupryianov, S. Bakurskiy, J. Jobst, T. Nakamura, K. Adachi, Y. Liu, and Y. Asano for valuable discussions and comments, and F. Hübler, Y. Nakamura, and Y. Yamaoka for their technical contribution. This work was supported by a Grant-in-Aid for Scientific Research on Innovative Areas "Topological Materials Science" (KAKENHI Grant Nos. JP15H05852, JP15K21717, JP15H05851), JSPS-EPSRC Core-to-Core program (A. Advanced Research Network), JSPS research fellow (KAKENHI Grant No. JP16J10404), Grant-in-Aid JSPS KAKENHI JP26287078 and JP17H04848, and the Netherlands Organisation for Scientific Research (NWO/OCW), as part of the Frontiers of Nanoscience program. VFB acknowledges support from the Foundation for Polish Science through the IRA Programme co-financed by EU within SG OP.

## Competing financial interests

The Authors declare no Competing Financial or Non-Financial Interests.

## Author contributions

The crystals were grown in the group of YM at Kyoto University. KL, YY, RF, and TT FIB-structured the crystals that were prepared by YY. YY, KL, and RF performed the transport measurements. MSA took part in the discussion. SY and YM supervised the measurements. YY and KL analyzed the results. VFB and MVM carried out the TDGL simulations. YY, KL, VFB, SY, MVM, JA, and YM wrote the paper with inputs from all the authors. YY and KL contributed equally to this work.

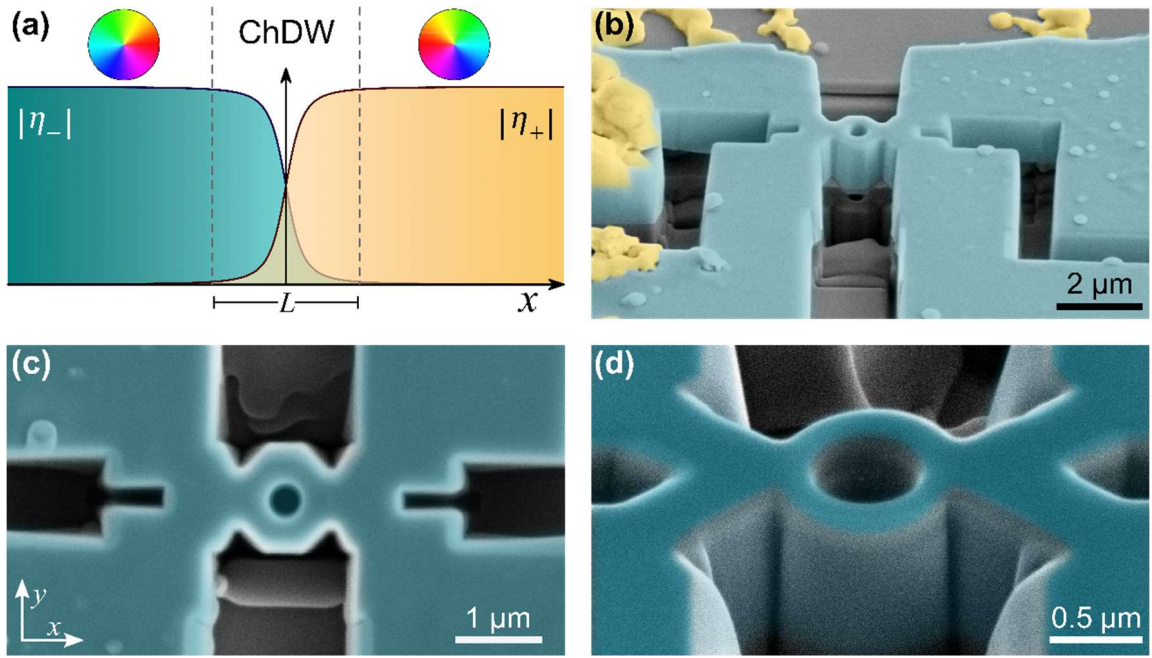

**Figure 1| Superconducting domain wall and the Sr₂RuO₄ microring.** (a) Schematic of a chiral domain wall (ChDW). $\eta_-$ and $\eta_+$ represent the degenerate chiral states meeting at a ChDW. The colour wheels represent the orbital phase of the chiral components, which wind in opposite directions. The two chiral wavefunctions overlap over a finite length $L$. As they locally suppress each other a Josephson junction is formed. (b) False-colour scanning electron microscope (SEM) image of Ring A. The blue represents the Sr₂RuO₄ crystal, and the yellow represents silver paint used for making electrical contact. Close-up images of (c) Ring A and (d) Ring B. In both rings the outer radius is around 0.55 µm while the inner radius of Ring B ($r_{in} = 0.3$ µm) is slightly larger than that of Ring A $r_{in} = 0.23 \pm 0.04$ µm. Each ring is connected to four transport leads and is sculpted out of a single crystal (around 0.7 µm thick) by a Ga⁺ focused ion beam.

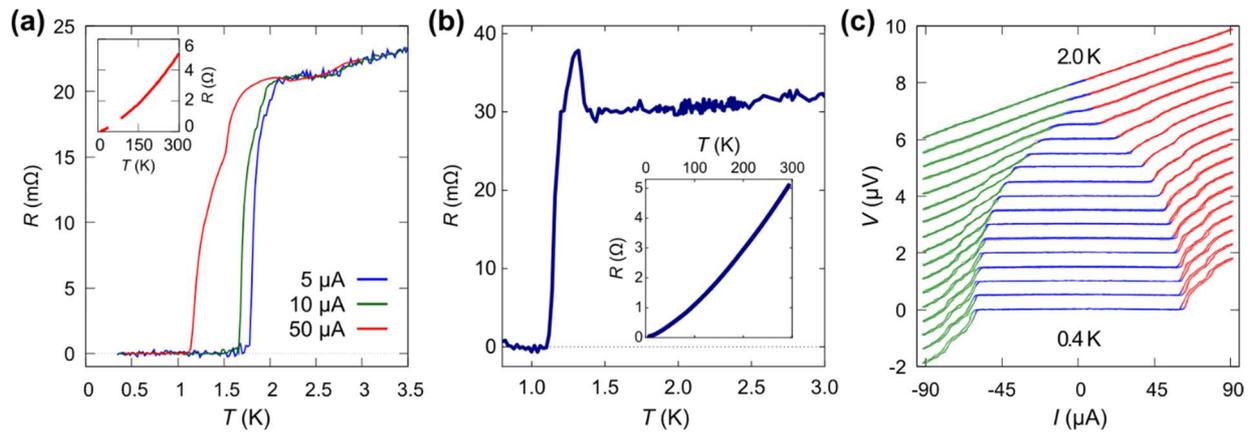

**Figure 2 | Basic properties of the Sr$_2$RuO$_4$ microrings.** Resistance as a function of temperature $R(T)$ for (a) Ring A and (b) Ring B. (a) presents data for various measurement currents, and (b) was measured using 10 μA. The insets show the $R(T)$ over a wider temperature range. Both rings exhibit highly metallic behaviour with residual resistivity ratio of 238 for Ring A and 177 for Ring B. (c) Current-voltage characteristics $V(I)$ of Ring A at various temperatures. The colours represent different voltage regions: $V < -0.1$ μV (green), $-0.1 < V < 0.1$ μV (blue), and $0.1$ μV $< V$ (red).

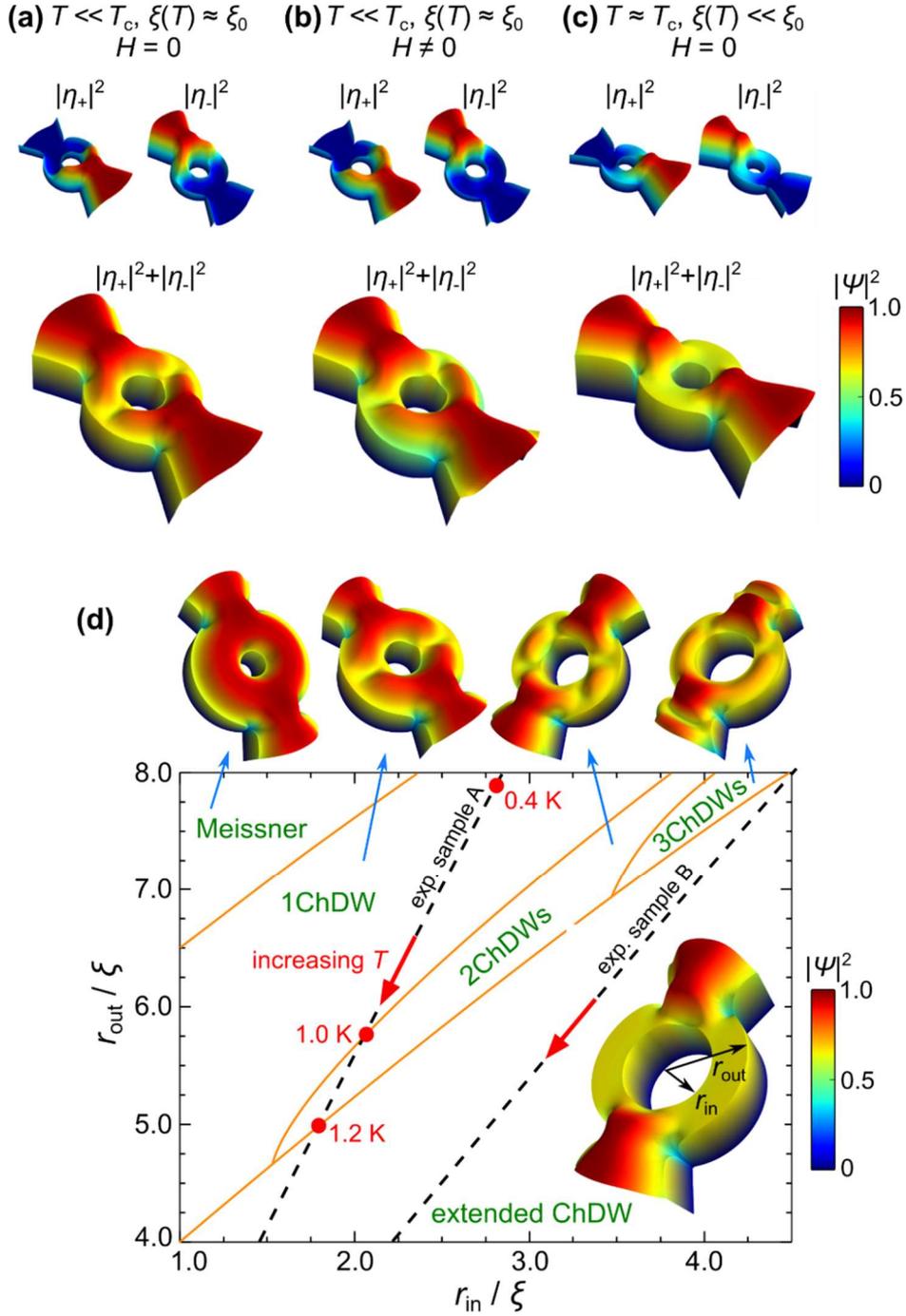

**Figure 3 | Simulated least-energy-state configurations of chiral *p*-wave microrings with nanostructured transport leads.** (a) State at a temperature much below $T_c$ without magnetic field (specifically, $T \approx 0.78$ K, $r_{in} = 2.5\xi$, and $r_{out} = 6.8\xi$). (b) State at the same temperature as (a) but with axial magnetic field. (c) State at a temperature close to $T_c$ in zero magnetic field (specifically, $T \approx 1.45$ K, $r_{in} = 1.3\xi$, and $r_{out} = 3.6\xi$). These states contain ChDWs that act as weak links of a SQUID. The colour maps represent the Cooper-pair density $|\Psi|^2$. In the panels (a) – (c), the inner $r_{in}$ and outer $r_{out}$ radii correspond to those of Rings A and B. The upper halves of the panels show the Cooper-pair density for each chiral component

$|\eta_\pm|$ of the corresponding states, for which the overall order parameter is expressed by $\Psi = \eta_+ k_+ + \eta_- k_-$. (d) $\left(\frac{r_{in}}{\xi(T)}, \frac{r_{out}}{\xi(T)}\right)$ Phase diagram in the absence of a magnetic field. For a wider ring arm, the "Meissner" state without ChDW is more stable; for a narrower ring arm, the extended ChDW state is expected. The dashed line shows the evolution of the least-energy states with increasing temperature according to the actual parameters of Ring A.

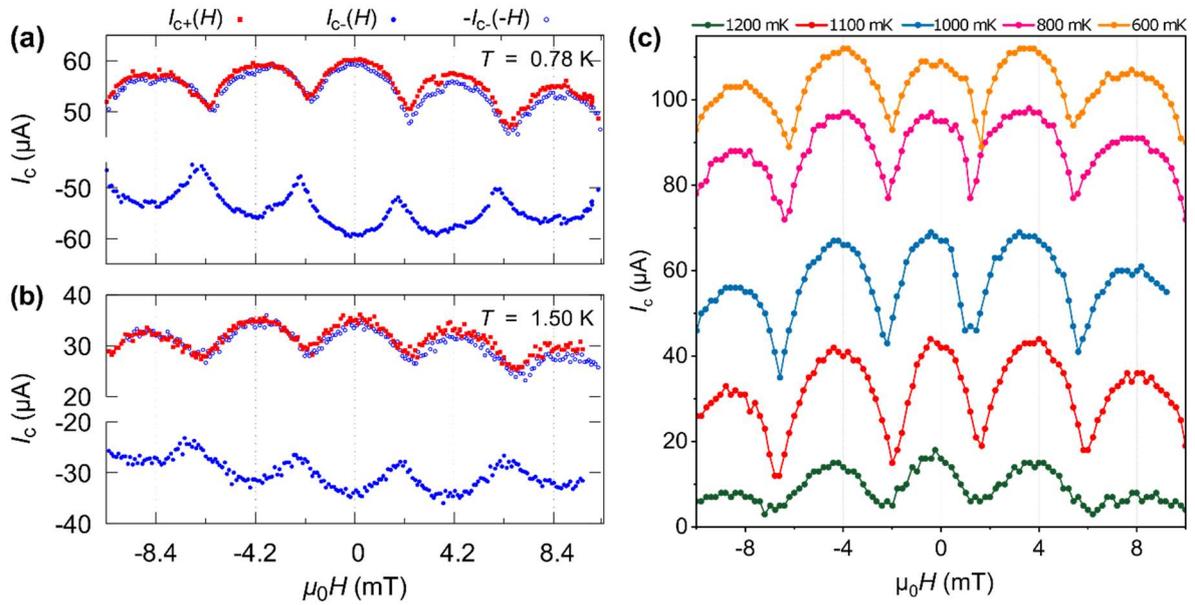

**Figure 4 | SQUID oscillations observed in the Sr$_2$RuO$_4$ microrings.** (a) Critical current as a function of magnetic field $I_c(H)$ of Ring A measured at 0.78 K and 1.50 K. The open blue circles show time-reversed critical current. (c) $I_c$ oscillations in Ring B over a wide range of temperatures. The $I_c$ values were obtained from $I(V)$ measurements at each magnetic field. The rings were heated up to above 5 K between each magnetic field.

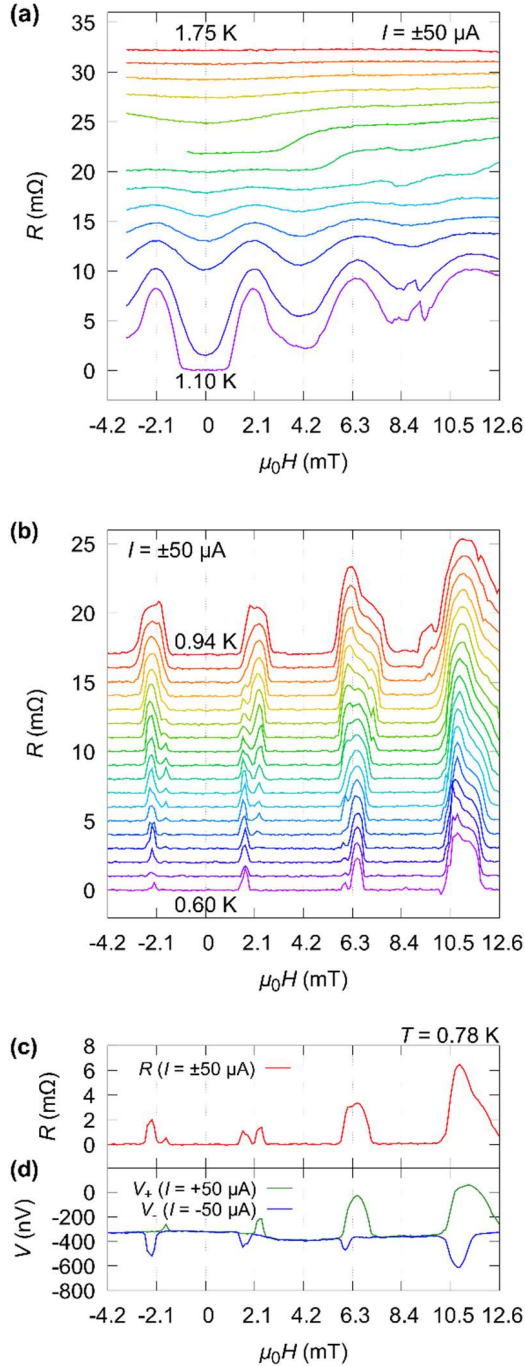

**Figure 5 | Magnetotransport of Ring A.** Magnetoresistance $R(H)$ with measurement current 50 μA for (a) temperatures between 1.75 K and 1.10 K and (b) temperatures between 0.94 K and 0.60 K. (c) $R(H)$ at 0.78 K. The $I_c$ oscillations at this temperature are displayed in Fig 4a. (d) The corresponding magnetovoltage when the measurement current is applied to one direction $V_+$ and to the other direction $V_-$. The peaks (dips) in $V_+$ ($V_-$) appear at different field values and hence double peaks appear in the resistance.

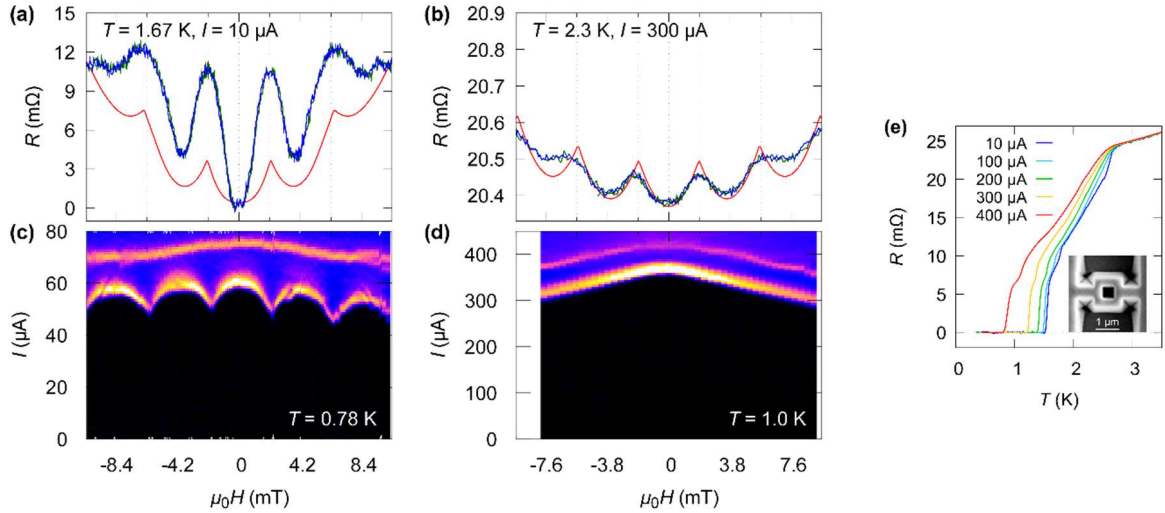

**Figure 6 | Comparison with a ring that exhibits the Little-Parks (LP) oscillations.** (a) Magnetoresistance $R(H)$ of Ring A near the transition temperature. (b) $R(H)$ of Ring C. The amplitude of the observed oscillations (blue and green) in Ring A is much greater than the expectation for the LP oscillations (red). In contrast, the observed oscillations in Ring C are in good agreement with a simulation for the LP oscillations. Colourmaps of the differential resistance $dV/dI$ of (c) Ring A and (d) Ring C as functions of magnetic field and measurement current. The bright part corresponds the critical current. The critical current of Ring C does not show any oscillation. (e) Resistance as a function of temperature of Ring C. The onset $T_c$ is 2.7 K, and hence the ring is in the extrinsic phase, which we consider as a non-chiral state. Data for panels (b) and (e) are adopted from [20].

Supplementary Information for

# Spontaneous emergence of Josephson junctions
# in homogeneous rings of single-crystal $Sr_2RuO_4$

## Supplementary Note 1.
## EFFECT OF RING STRUCTURING ON RING A

Here, we compare the effect of ring structuring to its superconductivity. Before the ring structuring, we milled the silver paint with focused ion beam (FIB) to make the four-wire configuration (Supplementary Figure 1a inset). Supplementary Figure 1 shows the resistance before/after the ring structuring. The transition temperatures $T_c$, where the resistance become zero, are not affected by the ring structuring. We note the zero-resistance $T_c$ of (a) is already slightly higher than 1.5 K.

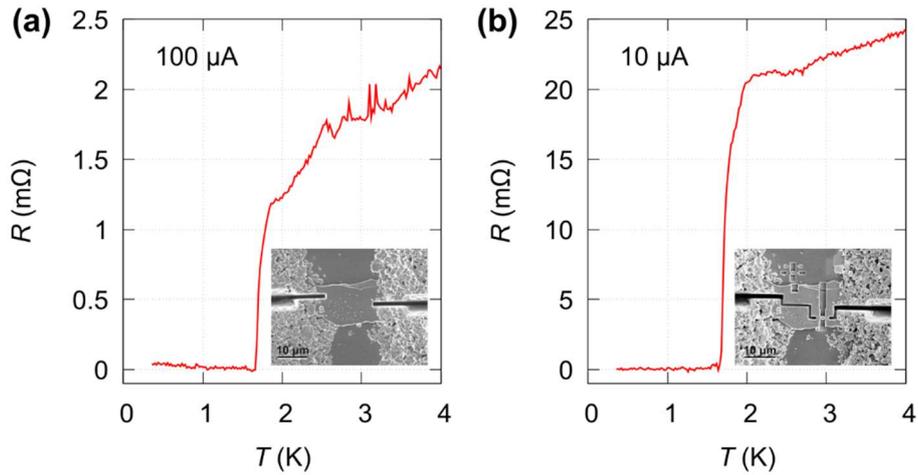

Supplementary Figure 1. Resistance as a function of temperature $R(T)$. (a) $R(T)$ before and (b) after ring structuring (Ring A). The zero-resistance $T_c$ is unchanged by the ring structuring.

## Supplementary Note 2.
## THEORY
### A. Theoretical formalism

The order parameter according to the irreducible representation that breaks the time-reversal symmetry in the crystallographic group of strontium ruthenate ($C_{4v}$) can be written in the form

$$\Delta(r, k) = \eta_+(r)k_+ + \eta_-(r)k_-, \qquad (S2)$$

where $k_\pm = (k_x \pm ik_y)/k_F$ are chiral basis functions in two dimensions, $k_F$ is the Fermi wave vector, and $\eta_\pm(r)$ are the space-dependent components of the order parameter. The differential equations satisfied by the components $\eta_\pm(r)$ are obtained from the minimization of the Ginzburg-Landau functional of $p$-wave superconductivity [1]:

$$\mathcal{F} = \alpha(|\eta_+|^2 + |\eta_-|^2) + \beta_1(|\eta_+|^2 + |\eta_-|^2)^2 - \beta_2(|\eta_+|^2 - |\eta_-|^2)^2$$
$$+ \frac{k_0 + k_1}{2}\{|\Pi\eta_+|^2 + |\Pi\eta_-|^2\}$$
$$+ (k_2 + k_3)\{(\Pi_-\eta_+)^*\Pi_+\eta_- + (\Pi_+\eta_-)^*\Pi_-\eta_+\} + \frac{B^2}{8\pi}, \qquad (S3)$$

where in the convenient notation $\hbar = c = k_B = 1$, microscopic calculations for a cylindrical Fermi surface (FS) yield $\alpha = -N(0)\ln(T_c/T)/2$, $\beta_1 = 21N(0)\zeta(3)/(8\pi T)^2$, $\beta_2 = \beta_1/3$, $k_1 = k_2 = k_3 = 7N(0)\zeta(3)v_F^2/2(8\pi T)^2$, and $k_0 = 3k_1$, with $N(0)$ being the density of states at the FS, $T_c$ the superconducting critical temperature, $\zeta(x)$ the Riemann zeta function, and $v_F$ the Fermi velocity [1]. Moreover, $\mathbf{\Pi} = (\mathbf{\nabla} - ie\mathbf{A})$ and $\Pi_\pm = (\Pi_x \pm i\Pi_y)/\sqrt{2}$.

The GL equations are obtained by minimization of the GL functional Eq. (S2), and are solved numerically in dimensionless form (see Ref. [2]), where all distances are scaled to the superconducting coherence length $\xi(T) = 21\zeta(3)v_F^2/(8\pi T)^2 \ln(T_c/T)$, the applied magnetic field to the bulk upper critical field $H_{c2} = \Phi_0/2\pi\xi^2$ ($\Phi_0$ is the flux quantum), and the Cooper-pair density to the bulk superconducting gap $\Delta_0 = (8\pi T)^2 \ln(T_c/T)/56\zeta(3)$. For convenience in the dimensionless quantities $\boldsymbol{\eta}' = (\eta'_+, \eta'_-)^T$, $\mathbf{\Pi}'$, $\Pi'_\pm$, $\mathbf{\nabla}'$, and $\mathbf{B}'$, from now on we drop all the primes to write the GL equations as

$$\frac{2}{3}\begin{bmatrix} \Pi^2 & \Pi_+^2 \\ \Pi_-^2 & \Pi^2 \end{bmatrix}\boldsymbol{\eta} + \left[1 - f(\beta)|\boldsymbol{\eta}|^2 + g(\beta)(\boldsymbol{\eta}^*\hat{\sigma}_z\boldsymbol{\eta})\hat{\sigma}_z\right]\boldsymbol{\eta} = 0, \qquad (S4)$$

$$\kappa^2 \nabla \times \nabla \times \mathbf{A} = \mathbf{J}, \qquad (S5)$$

where $\hat{\sigma}_z$ is the z Pauli matrix, $f(\beta) = (1-\beta)^{-1}$ and $g(\beta) = \beta(1-\beta)^{-1}$ are parameter functions depending on $\beta = \beta_2/\beta_1 = 1/3$, $\mathbf{J}$ is the superconducting current density, and $\kappa = \lambda/\xi$ is the GL parameter, with $\lambda$ being the London penetration depth. We take for κ the in-plane value of bulk SRO ($\kappa = 2.6$). Eq. (S4) is solved in 3D, taking the sample thickness into account following the Fourier approach described in Ref. [3].

The boundary conditions imposed on the components of the order parameter assumes specular reflection at the edges. They are obtained from the conditions that $\boldsymbol{\eta} \cdot \hat{\boldsymbol{n}} = 0$, and $\boldsymbol{J} \cdot \hat{\boldsymbol{n}} = 0$, where $\hat{\boldsymbol{n}}$ is a unitary vector normal to the boundary, and $\boldsymbol{J}$ is the superconducting current density. For the considered ring geometry of the sample, polar coordinate system is a convenient choice to solve the GL equations. Then, the supercurrent density is calculated by

$$J_\rho = \frac{2}{3}\text{Im}\left\{[\eta_+^*\Pi_\rho\eta_+ + \eta_-^*\Pi_\rho\eta_-] + \frac{1}{\sqrt{2}}[\eta_+^*e^{i\phi}\Pi_+\eta_- + \eta_-^*e^{-i\phi}\Pi_-\eta_+]\right\}, \quad (S6)$$

$$J_\phi = \frac{2}{3}\text{Im}\left\{[\eta_+^*\Pi_\phi\eta_+ + \eta_-^*\Pi_\phi\eta_-] + i\frac{1}{\sqrt{2}}[\eta_+^*e^{i\phi}\Pi_+\eta_- - \eta_-^*e^{-i\phi}\Pi_-\eta_+]\right\}. \quad (S7)$$

The GL equations (S3) and (S4) are solved in polar coordinates using finite differences on a uniformly spaced grid. Solutions to this equation are obtained by the iterative relaxation method, over time steps scaled to the Ginzburg-Landau time $\tau_{GL} = \xi^2/D$, where $D$ is diffusion constant. A large variety of initial inputs are provided to the algorithm to evaluate the stability of different solutions, and compare their free energy to find the ground state of the system.

The energy density (scaled to the condensation energy $F_0 = H_c^2 V/4\pi$, $V$ being the volume of the sample) used for the determination of the ground state is calculated after multiplication of Eq. (3) by $\eta^*$ from the left-hand side,

$$\frac{2}{3}(\eta^*\Pi^2\eta + \eta^*[\Pi_+^2\hat{\sigma}_+ + \Pi_-^2\hat{\sigma}_-]\eta) + |\eta|^2 - f(\beta)|\eta|^4 + g(\beta)(\eta^*\hat{\sigma}_z\eta)^2 = 0, \quad (S8)$$

where the matrices $\hat{\sigma}_\pm = (\hat{\sigma}_x \pm i\hat{\sigma}_y)/2$ have been introduced to simplify the notation. Straightforward calculations yield following expressions, useful to transform second-order derivatives into kinetic energy densities,

$$\eta^*\Pi^2\eta = \nabla \cdot (\eta^*\Pi\eta) - (\Pi\eta)^*\Pi\eta \quad (S9)$$

$$\eta^*[\Pi_+^2\hat{\sigma}_+ + \Pi_-^2\hat{\sigma}_-]\eta = \partial_+(\eta^*\hat{\sigma}_+\Pi_+\eta) + \partial_-(\eta^*\hat{\sigma}_-\Pi_-\eta) - \{(\Pi_-\eta_+)^*\Pi_+\eta_- + (\Pi_+\eta_-)^*\Pi_-\eta_+\}. \quad (S10)$$

The divergence term and the derivatives involving $\partial_\pm$ in Eqs. (S8) and (S9), where $\partial_\pm = (\partial_x \pm i\partial_y)/\sqrt{2}$, are discarded since they transform into vanishing surface terms. The substitution of the remaining terms into Eq. (S7) yields

$$\frac{2}{3}(|\Pi\eta_+|^2 + |\Pi\eta_-|^2 + \{(\Pi_-\eta_+)^*\Pi_+\eta_- + (\Pi_+\eta_-)^*\Pi_-\eta_+\}) = |\eta|^2 - f(\beta)|\eta|^4 + g(\beta)(\eta^*\hat{\sigma}_z\eta)^2, (S11)$$

which one can use to substitute the kinetic energy densities in the dimensionless form of Eq. (S2), and subsequently integrate to obtain the reduced free energy expression

$$F = -\frac{1}{2}\int(|\eta_+|^4 + |\eta_-|^4)dV - \frac{1+\beta}{(1-\beta)}\int|\eta_+|^2|\eta_-|^2dV + \kappa^2\int B^2 dV, \quad (S12)$$

where $V$ is the dimensionless volume of the sample.

The GL formalisms for chiral *p*-wave and chiral *d*-wave superconductors have analogous form, and the discussion here is applicable to both systems.

### B. Energetics behind domain-wall formation under mesoscopic confinement

For ring-shaped samples where the internal radius is kept fixed ($r_{\text{in}} = 2\xi$) and the external radius ($r_{\text{out}}$) is varied, the energy dependence on $r_{out}$ for a mono-domain state and a chiral domain-wall state (ChDW) is shown in Supplementary Figure 2. We also show the state with a circular domain-wall, that was predicted in Ref. [4] to be stable in mesoscopic samples, but exhibits much high energy in our case.

In Supplementary Figure 2 the state with one ChDW is the ground state of the system for $r_{\text{out}} < 7.7\xi$, and the mono-domain state becomes the ground state for $r_{\text{out}} > 7.7\xi$. Bearing in mind the temperature- and thickness- dependence of $\kappa$ in real samples, we considered a range of values from 1 to very large, but the threshold $r_{\text{out}}$ changed by not more than 10% (and we verified that for all the phases reported in Fig. 3d of the main manuscript). This small change in the threshold $r_{\text{out}}$ indicates the robustness of the reported ground states of the samples to magnetic screening (at least in absence of applied magnetic field).

To explain the mechanism behind the domain wall formation in what follows we analyze the contributions to the free energy of the system, using simplified expression of Eq. (S11). First, we note that the third term of Eq. (S11) has negligible (two orders of magnitude lower) values compared to the other two, in absence of applied magnetic field. Our calculated fields stemming from spontaneous currents, and their energy, are in accordance with earlier analysis of Ref. [5], for *p*-wave slabs. Although with weak contribution to total energy, the stray magnetic field of different states in Supplementary Figure 2 can serve as a mean for their identification in scanning-probe experiments. For that reason, we show in Supplementary Figure 3 the contour plots of the calculated distribution of Cooper-pair density (relevant to STM/STS) and the magnetic induction (relevant to MFM, SOT, SHPM) of those states.

To understand the physical significance of the remaining two terms in Eq. (S11), we show in Supplementary Figure 2 the contour plots of the energy densities corresponding to those terms, *i.e.* $\mathcal{F}_{\text{cond}} = |\eta_+|^4 + |\eta_-|^4$, and $\mathcal{F}_{\text{cross}} = |\eta_+|^2|\eta_-|^2$, for two sizes of the ring, namely $r_{\text{out}} = 6\xi$ and $r_{\text{out}} = 8.5\xi$, for both the mono-domain (conventional Meissner) and the ChDW state.

It is rather obvious that the energy density $\mathcal{F}_{\text{cond}}$ is maximal (minimal) wherever the Cooper-pair density is maximal (minimal). As a consequence, $\mathcal{F}_{\text{cond}}$ of the mono-domain state will always be lower than the one of the ChDW state, owing to the suppression of Cooper-pair density at the domain wall, since the suppression of Cooper-pair density (CPD) at the edges of the sample is nearly identical in the two states. In smaller samples, the $\Delta\mathcal{F}_{\text{cond}}$ contribution due to domain wall decreases as the suppression of CPD at the edges occupies more of the sample in both states, and the domain wall becomes shorter.

On the other hand, the energy density $\mathcal{F}_{\text{cross}}$ exhibits exactly opposite behavior to that of $\mathcal{F}_{\text{cond}}$. $\mathcal{F}_{\text{cross}}$ is maximal wherever two chiral components coexist, *i.e.* at the sample edges and along the ChDW. $\mathcal{F}_{\text{cross}}$ contribution increases as the sample is made smaller, but faster in the presence of a chiral domain wall, since then two chiral components (and associated spontaneous currents) become more spread over the sample.

To show this directly, in Supplementary Figure 2b we plot the energy differences $|\Delta F_{\text{cond}}| = |F_{\text{cond}}^{\text{ChDW}} - F_{\text{cond}}^{\text{monoD}}|$ and $|\Delta F_{\text{cross}}| = |F_{\text{cross}}^{\text{ChDW}} - F_{\text{cross}}^{\text{monoD}}|$ as a function of the sample size (outer radius). One can see that while $|\Delta F_{\text{cross}}|$ is the leading (largest) term in the energy for $r_{\text{out}} < 7.7\xi$, the condensation energy difference $|\Delta F_{\text{cross}}|$ dominates for $r_{\text{out}} > 7.7\xi$. In other words, the competition between these two energy terms directly determines the ground state of the system in absence of applied magnetic field.

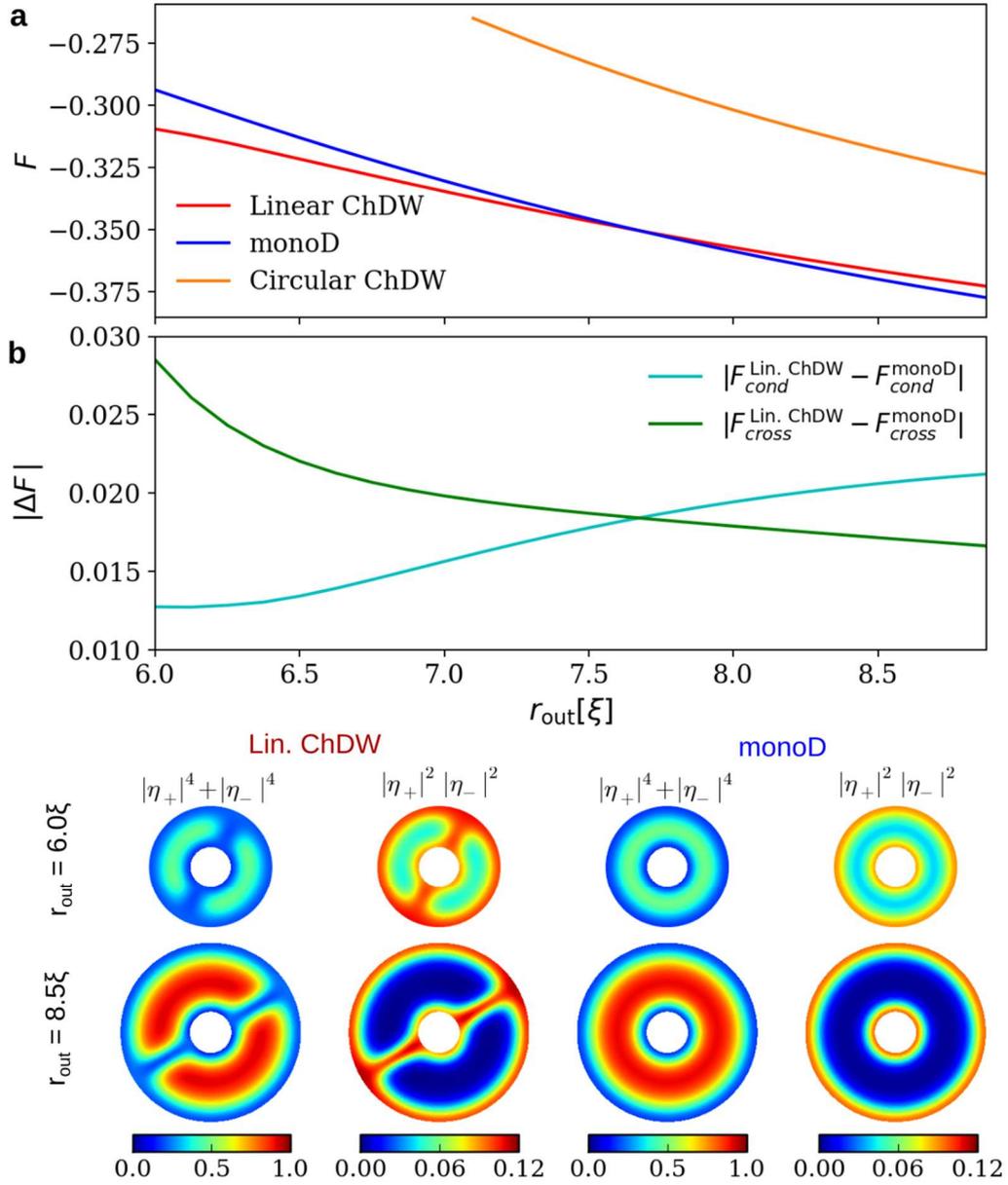

Supplementary Figure 2. (a) Energy dependence of the mono-domain (Meissner) state and the state with one chiral domain wall (ChDW) for ring samples with $r_{in} = 2\xi$ and varied $r_{out}$. For comparison, the energy of the state with a circular domain wall is also shown, albeit significantly higher. (b) Condensation and crossing energy difference ($\Delta F_{cond}$ and $\Delta F_{cross}$ respectively) between the mono-domain (Meissner) state and the ChDW state. Panel (c) shows the spatial distribution of $F_{cond}$ and $F_{cross}$, for samples with $r_{out} = 6.0\xi$ and $r_{out} = 8.5\xi$ in the Meissner state and the ChDW state.

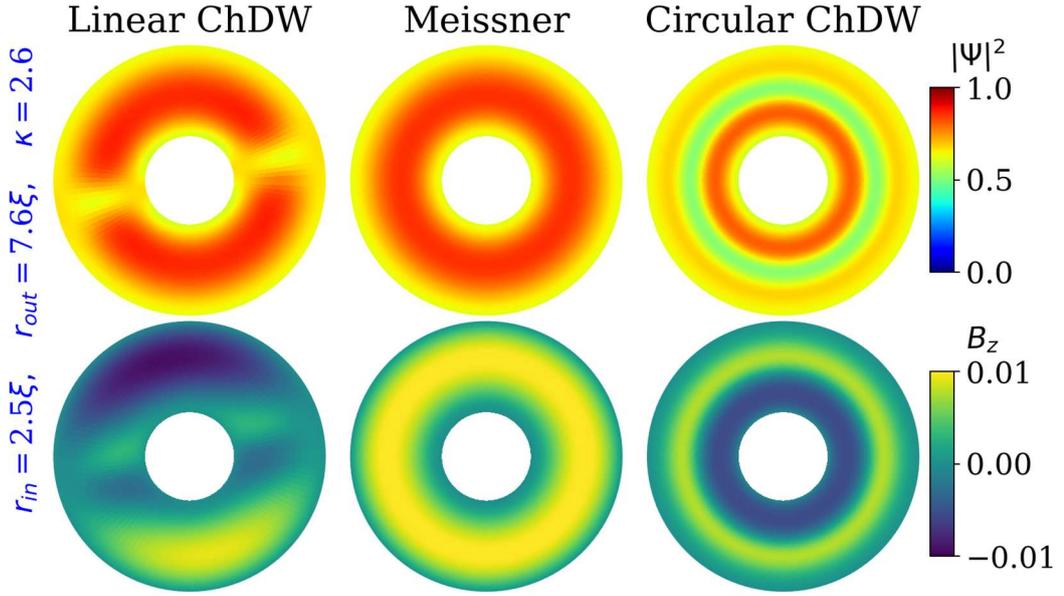

Supplementary Figure 3. Contour plots of the total Cooper-pair density and the magnetic induction for different states of Supplementary Figure 2a.

# Supplementary Note 3.
# ANOTHER EXAMPLE OF A RING IN THE EXTRINSIC PHASE

In this section, we show another ring (Ring D) in the extrinsic phase. Since this crystal was not covered with $SiO_2$ protection layer before FIB, the onset $T_c$ is higher than the other rings as shown in Supplementary Figure 4b, namely the ring is in the extrinsic (3-K) phase. Supplementary Figure 4c shows magnetoresistance $R(H)$ just above the zero-resistance $T_c$. The amplitude is not as large as that of a SQUID ring (Ring A). Therefore, the magnetoresistance oscillations of Ring D are considered to be the Little-Parks oscillations. We note the peaks appear non-periodically because the arm width $w = r_{\text{out}} - r_{\text{in}}$ is much wider than the coherence length, and hence the ring is outside the London limit [6].

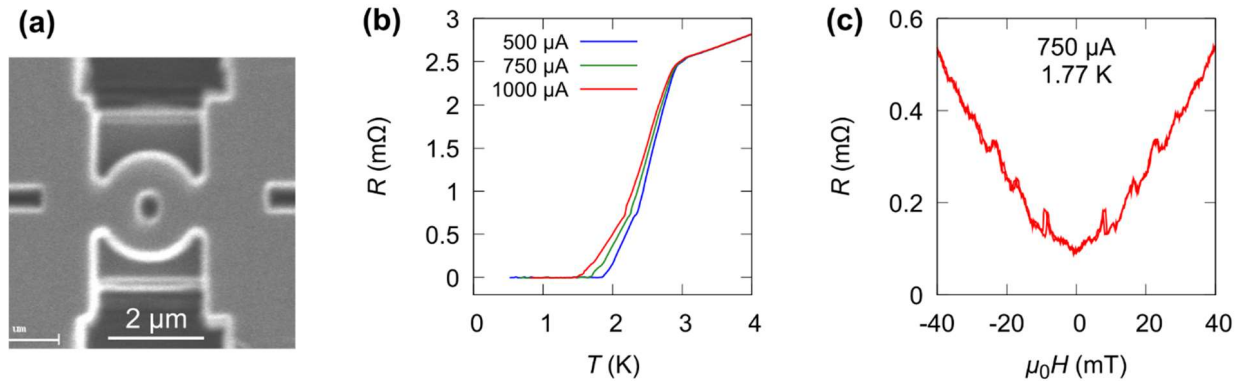

Supplementary Figure 4. Results of Ring D. (a) Scanning ion microscope image. (b) Resistance as a function of temperature $R(T)$. (c) Resistance as a function of magnetic field $R(H)$ at a temperature in the transition region. The amplitude of the magnetoresistance oscillations is much smaller than those of the rings without the extrinsic phase e.g. Ring A (see Fig. 5 in the main text).

# Supplementary Note 4.
# EVIDENCE OF DOMAIN WALL RECONFIGURATION

As shown in Fig. 3 of the main text, the Josephson junctions formed at the domain wall can take on a number of stable arrangements, depending on the size of $r_{in}$ and $r_{out}$ relative to $\xi(T)$. Most of these states result in the formation of parallel Josephson junctions in the arms of the ring, which would appear as SQUID oscillations in our transport measurements. Such is the case for Rings A and B, where any one of the simulated states for their given dimensions (1ChDW, 2ChDW and the extended ChDW depending on temperature) would yield similar SQUID oscillations, where the states cannot be distinguished from one another.

To find evidence of alternative domain configurations, we explore left side of the calculated phase diagram in Fig. 3, using a ring with considerably wider arms (Ring E). The ring dimensions ($r_{in} = 0.15$ μm, $r_{out} = 0.54$ μm) are chosen so that a ChDW would be energetically stable for an appreciable temperature range below $T_c$, but becomes progressively less favourable at lower temperatures as we approach the Meissner state. The results are summarised in Supplementary Figure 5.

Once again, we observe clear critical current oscillations over a wide temperature range below $T_c$. However, the supercurrent interference patterns differ significantly from those of Rings A and B. The distinction is perhaps most notable at $T = 1.35$ K, where the centre lobe of the $I_c(B)$ pattern is twice as wide as the side lobes, and the oscillation amplitude declines with approximately $1/B$ dependence. As shown in Supplementary Figure 5d, this $I_c(B)$ shows unambiguously the Fraunhofer pattern, which is considered as the hallmark of the Josephson effect. The observed Fraunhofer pattern serves to establish the presence of intrinsic junctions in our crystal, and hence supporting the existence of superconducting domain walls in $Sr_2RuO_4$. More importantly, the contrast between the Fraunhofer diffraction and the two-channel interference patterns of Rings A and B demonstrates a clear difference in the arrangement of the junctions (i.e. domain walls) based on ring dimensions.

Since a doubly connected superconductor cannot yield a Fraunhofer diffraction by itself (not even if one of the arms contains a weak link), the junction(s) needs to be arranged in series with the loop i.e. at either side of the ring, where it joins the transport leads via the (≈ 0.45 μm wide) "necks". While the walls of Rings A and B are considerably narrower than the leads on each side, arms of Ring E and the leads connected to it are comparable in width. It is therefore reasonable for domain walls to emerge from the constrictions on the sides of the ring rather than its centre. This also suggests that the domain walls are not pinned as strongly as those for Rings A and B. While this particular domain configuration is absent in the phase diagram of Fig. 3, we attribute this to the finite size of our simulations which do not capture the full effect of the large sections of the crystal connected to each side of the ring.

Also noteworthy is the evolution of the $I_c(B)$ pattern as we lower the temperature. Supplementary Figure 5c shows a substantial dampening of the oscillations as we approach the mono-domain (Meissner) state in Fig. 3. Note that even at $T = 400$ mK, where $\frac{r_{in}}{\xi(T)} \approx 2$ and $\frac{r_{out}}{\xi(T)} \approx 7$, the equilibrium Meissner state still has a slightly higher energy than one with a ChDW. However, the external magnetic fields applied while measuring $I_c(B)$ could tip this balance by pushing the domain wall outside our system.

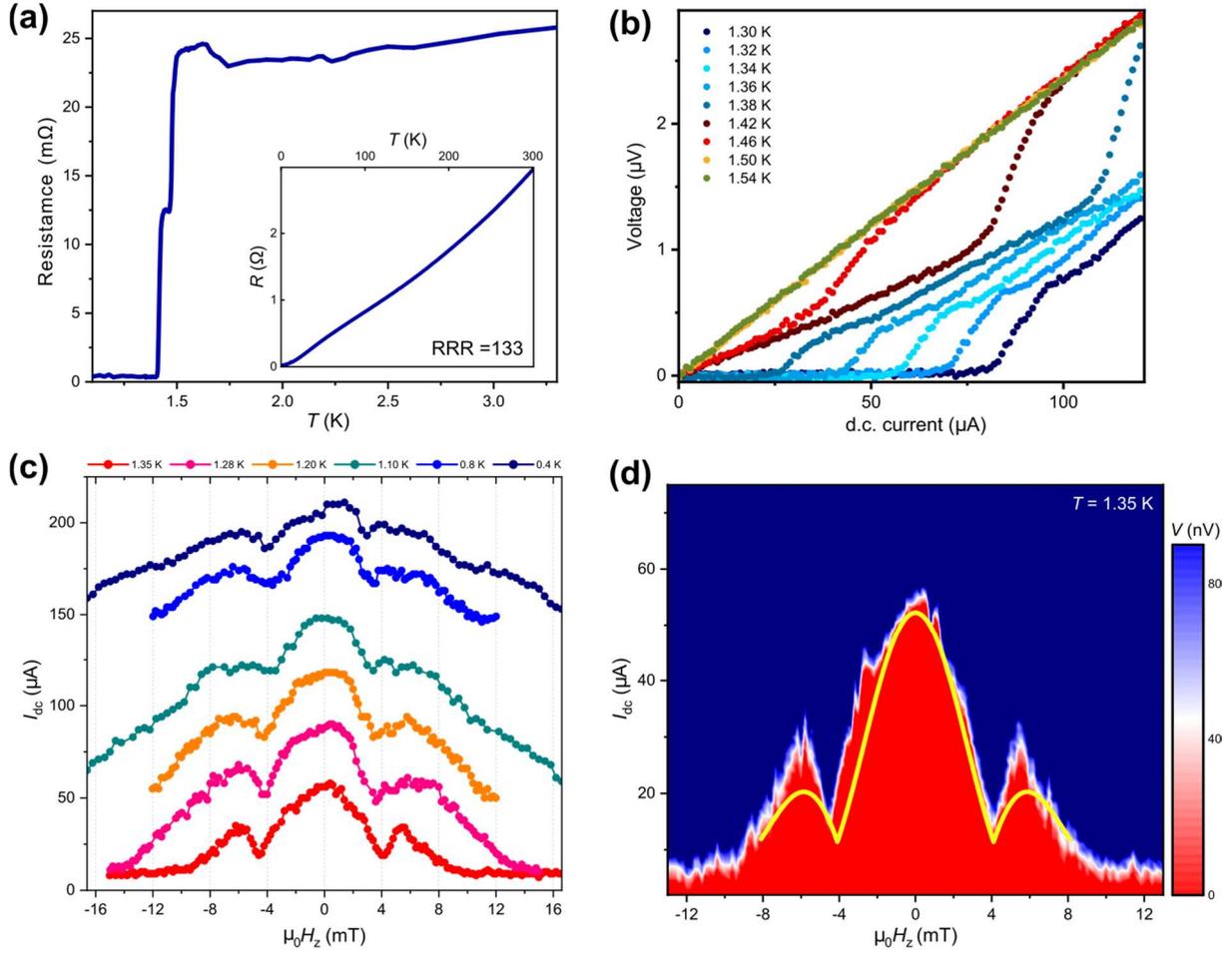

Supplementary Figure 5. Josephson effect in a ring with wider arms. (a) Resistance as a function of temperature $R(T)$ for Ring E ($r_{in} = 0.15$ μm, $r_{out} = 0.54$ μm), measured with a 10 μA bias. The large residual resistivity ratio (RRR = 133) shows the high metallicity of the system. (b) Current-voltage measurements taken at different temperatures. Superconductivity emerges at 1.47 K, close to the $T_c$ of bulk $Sr_2RuO_4$. (a) and (b) both indicate the complete absent of the extrinsic 3 K-phase. (c) Critical current as a function of out-of-plane magnetic field, measured at different temperatures. A Fraunhofer pattern is observed near $T_c$, where the center lobe (8.5 mT) is twice as wide as the side lobes (≈ 4 mT). This is shown more clearly in (d), where the data taken at $T = 1.35$ K is fitted with a Fraunhofer diffraction pattern. As the temperature is lowered to 0.4 K, ($\frac{r_{in}}{\xi(T)} \approx 2, \frac{r_{out}}{\xi(T)} \approx 7$) domain wall states become energetically less favourable in our system (see Supplementary Figure 2). Consequently, the $I_c(B)$ oscillations – driven by the Josephson effect at the domain wall – begin to fade away.

## Supplementary References